\documentclass[sigconf, screen, nonacm]{acmart}

\usepackage{textcomp}
\usepackage{multirow}
\usepackage{enumitem}
\usepackage{tikz}
\usepackage[capitalise,nameinlink]{cleveref}
\crefname{figure}{Fig.}{Figs.}
\Crefname{figure}{Fig.}{Figs.}
\crefname{section}{Sec.}{Secs.}
\Crefname{section}{Sec.}{Secs.}
\crefname{table}{Table}{Tables}

\ifdefined\pdfsuppressptexinfo\pdfsuppressptexinfo=-1\fi
\ifdefined\pdfsuppresswarningpagegroup\pdfsuppresswarningpagegroup=1\fi
\usepackage{array}
\newcolumntype{L}[1]{>{\raggedright\arraybackslash}p{#1}}
\AtBeginEnvironment{thebibliography}{\interlinepenalty=10000}

\begin{document}

\title{Compute Before You Digitize: ADC-less Mixed-Signal Processing for Scalable Cryogenic Qubit Readout}

\settopmatter{authorsperrow=3}
\author{Qirui Zhang}
\affiliation{\institution{University of Michigan}\city{Ann Arbor}\state{MI}\country{USA}}
\email{qiruizh@umich.edu}
\author{Ben Toby}
\affiliation{\institution{University of Michigan}\city{Ann Arbor}\state{MI}\country{USA}}
\email{btoby@umich.edu}
\author{He Wang}
\affiliation{\institution{University of Michigan}\city{Ann Arbor}\state{MI}\country{USA}}
\email{hewong@umich.edu}
\author{Benjamin Lienhard}
\affiliation{\institution{Technical University of Munich}\institution{Walther-Mei{\ss}ner-Institut}\city{Munich}\country{Germany}}
\email{benjamin.lienhard@tum.de}
\author{David Blaauw}
\affiliation{\institution{University of Michigan}\city{Ann Arbor}\state{MI}\country{USA}}
\email{blaauw@umich.edu}
\author{Dennis Sylvester}
\affiliation{\institution{University of Michigan}\city{Ann Arbor}\state{MI}\country{USA}}
\email{dmcs@umich.edu}
\renewcommand{\shortauthors}{Q. Zhang, B. Toby, H. Wang, B. Lienhard, D. Blaauw and D. Sylvester}

\begin{abstract}
Fault-tolerant quantum computing measures every ancilla qubit in every error-correction round, where readout is the slowest and least accurate operation. Scaling to thousands of qubits moves control and readout electronics into the cryostat, where readout consumes the largest share of a few-watt 4\,K budget. We show that a conventional readout RX baseband, with ADCs and a state-of-the-art machine-learning discriminator in carefully quantized fixed-point (FxP) logic, still takes 50\% of the 4\,K power per physical qubit, 85\% of it in Nyquist-rate ADCs and per-sample demodulation of a sparse, frequency-multiplexed spectrum. The 4\,K system thus draws merely 37\% less power than flex cables connecting room-temperature electronics, defeating the purpose of cryo-CMOS.

We introduce DIRAC, an ADC-less in-cryostat qubit readout architecture that uses analog computing to redraw the analog--digital boundary. Directly operating on the front-end's analog output, passive complex mixers perform the equivalent of digital demodulation, ping-pong active-RC integrators average their outputs into down-sampled per-qubit traces, and a time-domain stage digitizes them, eliminating Nyquist-rate ADCs. A TX--RX co-design derives the mixers' local oscillators from the readout transmitter's baseband rather than per-tone PLLs. A per-line calibration flow trains the matched filter jointly with the neural network, improving fidelity and enabling low-bit-width, low-voltage digital processing. Designed in 22\,nm FD-SOI and evaluated at 4\,K on 1.12\,M experimental shots, DIRAC matches the fidelity of a floating-point (FP32) digital pipeline for readout windows down to 750\,ns, at 31.6$\times$ and 238$\times$ lower energy than the FxP and FP32 pipelines. It cuts the total 4\,K power per physical qubit by 50.4\% and 68.6\% relative to FxP and room-temperature flex cables, respectively. At a fixed logical-qubit count and 4\,K budget, this yields, against FxP, a code distance larger by 2 to 6 and a 1.6--7.3$\times$ (Willow) and 4.9--204$\times$ (SI1000, $p = 10^{-3}$) lower logical error rate.
\end{abstract}

\maketitle

\vspace{-5pt}
\section{Introduction}\label{sec:intro}

\begin{figure}[t]
    \centering
    \includegraphics[width=\columnwidth]{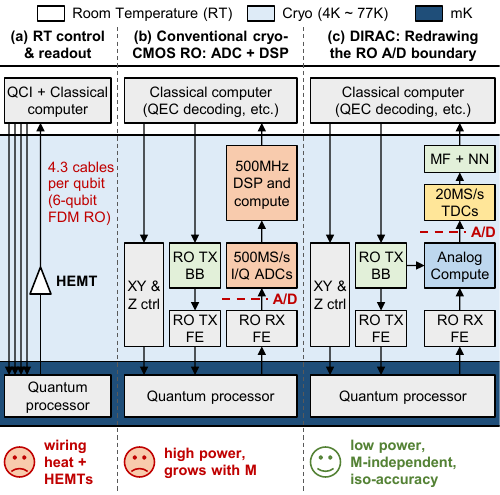}
    \Description{}
    \caption{Control and readout hardware solutions for large-scale cryogenic quantum computers: (a) all electronics at RT, (b) cryo-CMOS with conventional readout architecture, and (c) cryo-CMOS with DIRAC. RO: readout; TX/RX: transmitter/receiver; FE/BB: front-end/baseband.}
    \label{fig:teaser}
\end{figure}

Quantum computing promises transformative advantages in domains such as chemistry, materials and optimization, but practical problem instances require $10^5$ to $10^6$ physical qubits and logical error rates (LERs) as low as $10^{-15}$~\cite{beverland2022assessing,gidney2025how}. Physical qubits fail at rates around $10^{-3}$, so quantum error correction (QEC) encodes each logical qubit (LQ) redundantly in many physical qubits and repeatedly measures ancilla qubits to detect errors; superconducting qubits have recently operated a surface code below threshold~\cite{google2025quantum}. Reaching this scale breaks today's system architecture. Current superconducting machines connect room-temperature (RT) electronics to millikelvin qubits through long, lossy coaxial cables (\cref{fig:teaser} (a)), which does not scale because of (1) the intractable RT system and in-cryostat wiring complexity, (2) the heat these cables carry into the cryostat, and (3) long, variable feedback delays. RT electronics for a million transmons would occupy over 15,000\,m$^2$ and draw about 40\,MW~\cite{bardin2021microwaves}. QIsim~\cite{min2023qisim} finds that cable heat limits RT electronics to fewer than 700 qubits, whereas cryogenic CMOS (cryo-CMOS) interfaces (e.g., at 4\,K) avoid the cable heat and are limited only by their own power, which architecture and circuit improvements can reduce. Industry and academia have therefore embraced cryo-CMOS control and readout, from Google~\cite{bardin2019design,yoo2023design}, IBM~\cite{chakraborty2022cryo, chakraborty2024rfawg} and Intel~\cite{park2021fully} to academic groups~\cite{vandijk2020scalable,kang2022cryo, kang2025quantum,guo2023polar}, and industrial roadmaps plan for it~\cite{mohseni2024build}. Power, however, is scarce at 4\,K. A dilution refrigerator offers only about 1.5\,W at 4\,K~\cite{krinner2019engineering}, and even kilowatt-class helium plants leave about 1\,mW per qubit at the million-qubit scale~\cite{bardin2019design}. Cryo-CMOS pays off only if it draws far less power than the cables it replaces, while keeping the fidelity of RT electronics.

Qubit readout is currently the bottleneck of this system. First, every QEC round ends by measuring all ancilla qubits, and readout is among the least accurate operations: on Google's 72-qubit Willow processor, the mean measurement error is 0.8\%, against 0.35\% for a CZ gate and 0.062\% for a single-qubit gate~\cite{google2025quantum}. Second, readout typically takes more than half of a QEC round (500 of 921\,ns in Google's surface-code experiment~\cite{google2023suppressing}), and its hardware can take a large share of the 4\,K power: in QIsim's 4\,K CMOS interface, the digital part of the readout receiver (RX) alone accounts for 54.7\% of the total power~\cite{min2023qisim}. Machine-learning (ML) discriminators recover a large part of the readout error on frequency-division-multiplexed (FDM) setups, where several qubits share one readout line: fully connected neural networks (NNs)~\cite{lienhard2022deep,maurya2023herqules,mude2025efficient} and transformer and CNN discriminators~\cite{giortamis2026mcmit} improve fidelity substantially over conventional discrimination with matched filters (MFs) alone~\cite{lienhard2022deep}. All of them, however, are evaluated in software or on RT FPGAs, with no account of their cost inside a cryostat, and some models can be too large to run at 4\,K, such as a fully connected network with about 1.6\,M parameters~\cite{lienhard2022deep} or a transformer with hundreds of thousands of parameters~\cite{giortamis2026mcmit}. Cryo-CMOS readout chips, on the other hand, either focus only on the front-end (FE)~\cite{nagulu2023sub,vanwinckel2022cryo} or implement a baseband (BB) too simple to support FDM readout and ML discrimination at the same time~\cite{minn2025cryo,guo2024cryo,guo2026cryogenic,huang2025cryogenic,guo2026reflectometry}, although both are needed to improve system scalability and readout fidelity together.

In this paper, we envision a cryogenic quantum computer whose quantum-classical interface (QCI) sits at 4\,K and study the readout architecture it needs. To the best of our knowledge, this is the first study of a cryo-CMOS readout architecture that supports ML-based discrimination of frequency-multiplexed qubits. We start from a baseline that follows the conventional readout chain~\cite{lienhard2022deep,maurya2023herqules,mude2025efficient} of RT QCIs (\cref{fig:teaser} (b)): 500\,MS/s ADCs digitize the intermediate-frequency (IF) output of the RX FE, and a digital pipeline demodulates every tone at the same data rate and runs MF banks with per-qubit NNs, evaluated with a measured 4\,K ADC~\cite{lee2023programmable} and post-layout digital power. In FP32, this readout RX BB consumes 89\% of the 4\,K power per physical qubit and raises the total to 7.35\,mW, more than the 6.27\,mW of using coaxial cables to connect an RT QCI. Carefully quantized to fixed point (FxP) at iso-accuracy, it still requires 50\%, so the 4\,K system draws merely 36.7\% less power than using flex cables to connect an RT QCI, too small a margin to justify moving the QCI into the cryostat.

To address this bottleneck, we introduce DIRAC, an A\textbf{D}C-less \textbf{I}n-cryostat qubit \textbf{R}eadout architecture that uses \textbf{A}nalog \textbf{C}omputing to redraw the analog--digital (A/D) boundary (\cref{fig:teaser} (c)), drastically reducing readout power consumption. Our key observation is that the baseline spends power not on discrimination, but on the Nyquist-rate processing its A/D boundary imposes: the ADCs digitize the IF band and demodulators process samples at this high rate, only to produce per-qubit block averages at a much lower rate. Demodulation and block averaging are fixed, linear computations that analog circuits perform cheaply; only the small MF and NN must remain digital to be trainable. DIRAC splits the dataflow accordingly: an analog micro-architecture (mixers, integrators and time-domain digitizers) computes directly on the analog RX FE output and digitizes only the down-sampled block averages, so no high-cost Nyquist-rate ADC is left. This raises two problems, solved by co-design. First, analog demodulation needs local oscillators (LOs) for every tone; DIRAC derives them from the readout transmitter (TX) BB, not from per-tone PLLs. Second, analog circuits are imprecise and vary by line; a per-line calibration flow trains the MF jointly with the NN, absorbing these non-idealities and permitting low-bit-width digital processing at low voltages with forward body bias (FBB). Our contributions are as follows:

\begin{enumerate}[label=\arabic*), wide=0pt, topsep=2pt, itemsep=0pt, parsep=0pt]
\item \textbf{Baseline and Analysis} (\cref{sec:motivation}): We build FP32 and iso-accuracy FxP readout RX BBs for ML-based FDM readout, with post-layout 4\,K power in 22\,nm FD-SOI, as the first cryo-CMOS baselines of their kind. ADCs and digital demodulation take 85\% of the FxP energy, which motivates DIRAC.
\item \textbf{DIRAC Architecture} (\cref{sec:arch}): A novel qubit readout architecture that splits the dataflow at a redrawn A/D boundary: the fixed, linear stages (demodulation and block averaging) are computed in analog, and only their down-sampled outputs are digitized, without high-cost Nyquist-rate ADCs. The analog chain and its digitizers draw only 44.2\,\textmu{W} per qubit while measuring and use 61.2$\times$ and 486$\times$ less energy than the same stages in FxP and FP32.
\item \textbf{TX--RX Baseband Co-Design} (\cref{sec:arch-lo}): Typical readout TX BBs, however they generate the FDM tones, already provide the sign bits of every tone's cosine and sine. These are the BB LOs needed by DIRAC's mixers, so reusing them removes the need for per-tone PLLs ($\sim$80\,\textmu{W} each~\cite{huang2025cryogenic}).
\item \textbf{Learned Calibration and Co-Design} (\cref{sec:learning}): We propose a per-line calibration flow that trains the MF jointly with the NN on each line's own outputs to absorb analog non-idealities and variation, recovering +0.41\% of $F_{4Q}$. This allows MF and NN to be quantized aggressively, complemented by a set of architecture--circuit co-designs that run them in separate clock and supply domains at low voltages with FBB.
\item \textbf{Cryo-CMOS Implementation and Evaluation} (\cref{sec:method,sec:eval}): We design DIRAC in 22\,nm FD-SOI at 4\,K and evaluate its accuracy with a behavioral model cross-validated against transistor-level simulation. On 1.12\,M experimental shots over ten mismatch-sampled chips, it reaches $F_{4Q}$ = 95.38\,$\pm$\,0.02\% (FP32: 95.28\%), matches FP32 for readout windows down to 750\,ns, and spends 31.6$\times$ and 238$\times$ less energy than FxP and FP32. It cuts the 4\,K QCI power per physical qubit by 50.4\% and 68.6\% relative to FxP and RT flex cables, so a 4\,K budget holds 2.0$\times$ as many LQs as with FxP or, at a fixed LQ count, allows a code distance larger by 2 to 6 and a 1.6--7.3$\times$ (Willow) and 4.9--204$\times$ (SI1000) lower LER.
\end{enumerate}

\section{Background}\label{sec:background}

\begin{figure}[t]
    \centering
    \includegraphics[width=\columnwidth]{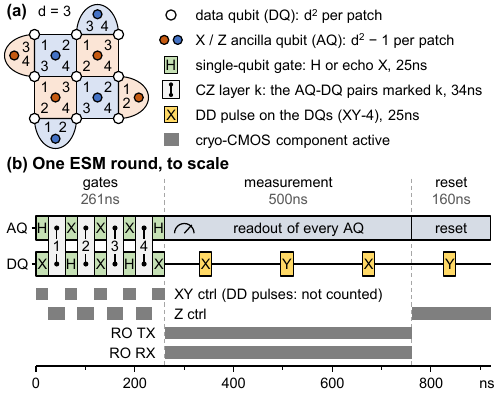}
    \Description{}
    \caption{The surface-code ESM round under Google's schedule~\cite{google2023suppressing}: (a) a distance-3 patch numbered by CZ layer, and (b) operations and durations, with the control and readout hardware active in each step.}
    \label{fig:esm}
\end{figure}

\begin{figure}[t]
    \centering
    \includegraphics[width=0.8\columnwidth]{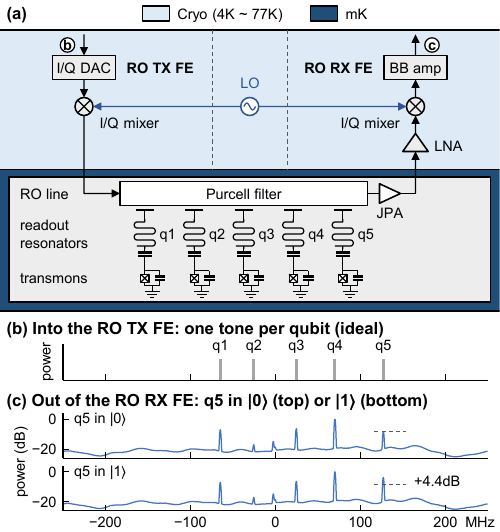}
    \Description{}
    \caption{FDM readout for superconducting qubits: (a) architectures of the quantum processor and cryo-CMOS FEs, (b) spectrum of the ideal TX BB signal, and (c) average spectra of the RX BB signal.}
    \label{fig:readout}
\end{figure}

\subsection{Surface Code Syndrome Measurement}\label{sec:background-qec}
\edef\DefaultTopNumber{\arabic{topnumber}}\setcounter{topnumber}{1}
We study the cryo-CMOS QCI, i.e., the control and readout electronics at 4\,K, of large-scale fault-tolerant quantum computers using the surface code. A distance-$d$ rotated surface code (\cref{fig:esm} (a)) stores one logical qubit in $d^2$ data qubits (DQs) and $d^2-1$ ancilla qubits (AQs)~\cite{google2023suppressing}. In each error-syndrome measurement (ESM) round, each AQ checks the parity of its two to four DQ neighbors, and all AQs are measured in parallel to produce syndromes for decoding. In Google's surface-code experiment (\cref{fig:esm} (b))~\cite{google2023suppressing}, a round consists of five layers of 25\,ns single-qubit gates interleaved with four layers of 34\,ns CZ gates, followed by a 500\,ns measurement and a 160\,ns reset of AQs, 921\,ns in total. The numbered tile corners in \cref{fig:esm} (a) give the CZ order: in layer $k$, each AQ interacts with the DQ at its corner marked $k$, so each layer pairs every qubit with a different neighbor. While the AQs are measured and reset, the DQs idle under dynamical decoupling (DD).

\subsection{Cryo-CMOS Quantum-Classical Interface}\label{sec:bg-cryo}
In a superconducting quantum computer, if all electronics stay at RT (\cref{fig:teaser} (a)), the 4\,K stage absorbs the heat of about 4.3 cables per qubit (\cref{sec:method-system}). QIsim~\cite{min2023qisim} shows that scalability then depends entirely on the heat load of these cables, whereas a 4\,K QCI is limited by its own power; as architecture, circuits, and technology improve, cryo-CMOS may therefore scale best. At 4\,K, it also offers higher density and more mature design tools than SFQ logic~\cite{knapen2026pinball}. We thus build on cryo-CMOS, which removes the cables but pays off only if its 4\,K power stays low.

\cref{fig:teaser} (b) shows a typical cryo-CMOS QCI. An XY controller per qubit drives single-qubit gates with RF pulses at the qubit frequency (4--8\,GHz)~\cite{bardin2019design,yoo2023design}, a Z controller per qubit or coupler drives flux bias pulses for CZ gates and reset~\cite{yoo2023design,google2023suppressing}, and the readout hardware, a TX~\cite{vanwinckel2022cryo} and an RX~\cite{nagulu2023sub}, measures the AQs (\cref{sec:bg-fdm}). The bottom row of \cref{fig:esm} (b) marks when each block is active: XY in the single-qubit layers, Z in the CZ layers and reset, and readout in the measurement, the longest step. Lattice surgery, the standard way to compute on surface-code logical qubits, merges and splits patches and measures each merged patch for $d$ ESM rounds~\cite{horsman2012surface,litinski2019game}, so ESM on all patches sets the peak power of the QCI~\cite{min2023qisim}.

\begin{figure*}[t]
    \centering
    \includegraphics[width=0.92\textwidth]{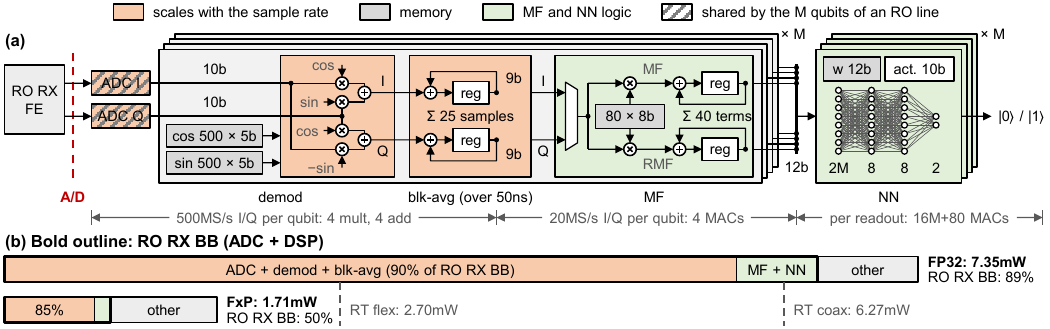}
    \Description{}
    \caption{Conventional readout RX architecture: (a) the digital BB of one qubit, with its bit widths after FxP quantization, and (b) the 4\,K power breakdown of the cryo-CMOS QCI with the FP32 and FxP readout RX BBs.}
    \label{fig:baseline}
\vspace{-8pt}
\end{figure*}

\subsection{Qubit Readout and State Discrimination}\label{sec:bg-fdm}\label{sec:bg-algo}
\cref{fig:readout} (a) shows the quantum processor and cryo-CMOS FEs of FDM readout~\cite{lienhard2022deep}: each qubit is measured via a readout resonator whose frequency depends on its state, and one readout line with one set of readout hardware serves $M$ qubits, whose tones are about 40--60\,MHz apart ($M$ = 6 in Google's processors~\cite{google2023suppressing}). The readout TX synthesizes and sums the per-qubit tones in its BB and up-converts the sum in its FE (\cref{fig:readout} (b)). Each resonator changes the amplitude and phase of its tone with its qubit's state; the RX FE amplifies the line and mixes it down to IF~\cite{nagulu2023sub}, and the RX BB discriminates every qubit's state. The IF spectrum is sparse: each tone is a narrow peak whose slowly varying envelope carries the state. \cref{fig:readout} (c) shows a measured example: averaged over all traces with the other qubits in $|0\rangle$, q5's tone is 4.4\,dB stronger in $|1\rangle$ than in $|0\rangle$, so the qubit states can be discriminated from the IF signal.

We use a measured five-qubit dataset~\cite{lienhard2022deep} and design and evaluate our system around the signals of q1, q3, q4 and q5, since prior work finds q2's states poorly distinguishable~\cite{lienhard2022deep,maurya2023herqules,mude2025efficient}; we thus report $F_{4Q}$, the geometric mean of their assignment fidelities~\cite{maurya2023herqules}. Without loss of generality, our discriminator builds on HERQULES~\cite{maurya2023herqules}, which, unlike fully connected NN~\cite{lienhard2022deep} and transformer~\cite{giortamis2026mcmit} discriminators, targets high fidelity at minimal hardware cost, the best fit for the low power that cryo-CMOS requires. Following its later version~\cite{mude2025efficient}, we split the shared NN into per-qubit heads, but keep binary labels since the dataset prepares no $|2\rangle$ states; \cref{sec:mot-pipeline} details the configuration. Any other discriminator that works on per-qubit demodulated, block-averaged traces can in principle run behind DIRAC, with its digital part extended accordingly.

\section{Motivation}\label{sec:motivation}
\setcounter{topnumber}{\DefaultTopNumber}

\subsection{Cryo-CMOS Baseline Designs}\label{sec:mot-pipeline}\label{sec:mot-fx}
A conventional readout RX BB consists of ADCs, which digitize the IF output of the RX FE, and a digital state discriminator. To quantify its power in cryo-CMOS, we design the discriminator and pair it with a published 4\,K ADC~\cite{lee2023programmable} (\cref{fig:baseline} (a)). An I/Q ADC pair, shared by the $M$ qubits of a readout line, digitizes the RX FE output at the Nyquist rate of the IF band, $f_s$ = 500\,MS/s. Each qubit then has its own pipeline. A complex demodulator multiplies every I/Q sample by the qubit's digital LO, read from cos/sin tables at its tone frequency, and a block-averaging stage (blk-avg) sums every 25 products into one block average (20 per 1\,\textmu{s} readout). A filter bank computes two scores per qubit, the MF and the relaxation MF (RMF)~\cite{maurya2023herqules}, each an inner product over the I and Q parts of the 20 block averages (40 terms). Finally, a 2$M$-8-8-2 NN head, one per qubit as in~\cite{mude2025efficient}, decides the qubit's state from the scores of all $M$ qubits, so that it learns the crosstalk among them.

We implement the discriminator in 22\,nm FD-SOI with a single 500\,MHz clock, a 0.8\,V supply, and fine-grained clock gating, in two number formats: FP32, the precision in which it is trained and evaluated in software~\cite{maurya2023herqules}, is the accuracy reference, and FxP uses the smallest bit widths that preserve this accuracy. The dataset was recorded by ADCs with 10.6b ENOB~\cite{lienhard2022deep,keysight2022m3102a}, so we model the ADCs with the 10b, 500\,MS/s mode of a cryo-CMOS ADC built specifically for FDM qubit readout~\cite{lee2023programmable}. We then quantize one stage at a time, with the stages behind it still in FP32: each stage takes the smallest width that keeps the validation $F_{4Q}$ within 0.005\% of the previous step, and the NN heads are retrained with quantization-aware training (QAT). The result is 5b cos/sin tables, 9b block averages, 8b MF coefficients, 12b MF scores, 12b NN weights and 10b activations (\cref{fig:baseline} (a)). On the test set, FxP matches the fidelity of FP32 (95.28\%), on par with HERQULES as evaluated by MCMit~\cite{giortamis2026mcmit,giortamis2026oraqle} (95.1\%) and thus a fair basis for our hardware study.

\subsection{Power Analysis and Key Observation}\label{sec:mot-power}
Both designs are synthesized, placed and routed with a standard-cell library characterized at 4\,K (\cref{sec:method-hw}). \cref{fig:baseline} (b) places them in the 4\,K power of the complete QCI over an ESM round (\cref{sec:method-system}). With FP32, the 4\,K QCI draws 7.35\,mW per physical qubit, of which the RX BB takes 89\%. With FxP, the total falls to 1.71\,mW, but the RX BB still takes 50\% of it. Both designs defeat the purpose of cryo-CMOS: the FP32 system draws more 4\,K power than the coaxial cables that connect an RT QCI (6.27\,mW per qubit), and the FxP system only 36.7\% less than flex cables (2.70\,mW), a margin too small to justify moving all QCI electronics into the cryostat.

Within the RX BB, the power is concentrated ahead of the MF. The ADC pair, the demodulators, and the block-averaging stages take 85\% of the FxP RX BB energy (FP32: 90\%; \cref{sec:eval-energy}), yet they only produce the block averages; the MF and the NN, which discriminate the state, take only the small remainder. These stages are inefficient because they process a sparse spectrum at the full Nyquist rate: the ADCs digitize the entire IF band at 500\,MS/s and 10b, and every qubit's demodulator processes all of these samples, although the information lies in $M$ narrow tones whose envelopes vary on a 50\,ns scale (\cref{fig:readout} (c)).

This architecture also scales poorly with $M$. More tones per line widen the IF band, so $f_s$ must rise in proportion to $M$. ADC power then grows super-linearly, since converter efficiency degrades at high sample rates~\cite{murmann2026adc}. The digital cost scales no better: each demodulator processes samples at $f_s$, so the cost per qubit, expected to be constant, grows with $M$, and that of a readout line grows quadratically instead of linearly.

\begin{figure*}[t]
    \centering
    \includegraphics[width=0.92\textwidth]{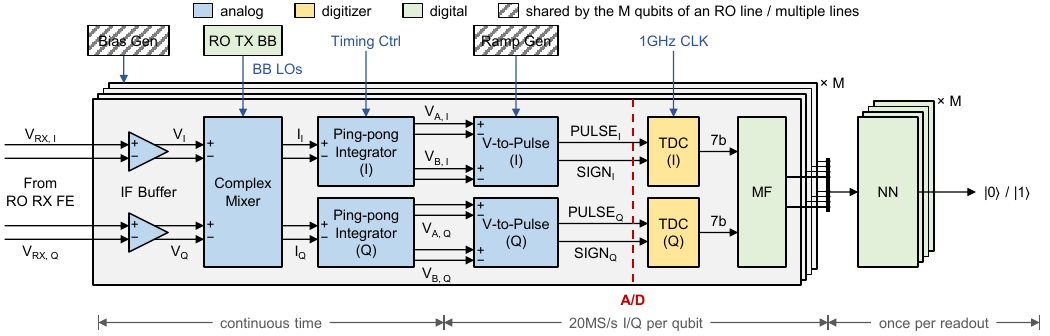}
    \Description{}
    \vspace{-4pt}
    \caption{DIRAC architecture for one qubit of a readout line.}
    \label{fig:arch}
\vspace{-8pt}
\end{figure*}

\textbf{Key observation.} The baseline spends its power not on discrimination, but on the Nyquist-rate processing that its A/D boundary imposes: since the IF band is digitized before demodulation, every stage ahead of the MF runs at $f_s$. Demodulation and block averaging, however, are fixed, linear computations. If they were performed before the A/D boundary, conversion would drop to one block average per qubit every 50\,ns, regardless of $f_s$ and $M$, and only the small, trainable MF and NN would remain digital. \cref{sec:arch} redraws the boundary accordingly.

\section{DIRAC Architecture}\label{sec:arch}
\subsection{Overview: Redrawing the A/D Boundary}\label{sec:arch-overview}\label{sec:arch-buf}
Motivated by this observation, DIRAC redraws the A/D boundary of cryogenic qubit readout, moving it behind demodulation and block averaging. It processes the RX FE's IF output directly in the analog domain (\cref{fig:arch}). For fan-out to the line's $M$ qubits, each qubit starts with two source-follower IF buffers (I/Q) with good input--output isolation, a gain of 0.92 and a 1.07\,GHz bandwidth, well above the 250\,MHz IF band, at 11.3\,\textmu{W} per qubit. A passive complex mixer with square-wave LOs demodulates the qubit's tone, and two ping-pong integrators each for I and Q compute its block average over every 50\,ns slot in continuous time. A V-to-Pulse stage turns each held value into a pulse of proportional width and a sign, and a time-to-digital converter (TDC) counts clock edges during the pulse. The two TDCs (I and Q), the only digitizers, form the A/D boundary: each outputs a sign and a 6b magnitude per block average, at the down-sampled rate $f_\mathrm{ds}$ = 20\,MS/s.

The energy of one readout of length $t_\mathrm{meas}$ makes the difference explicit. Per qubit, the baseline spends
\begin{equation}\label{eq:edig}
E_\mathrm{Baseline} = t_\mathrm{meas} \Big( \frac{2 f_s}{M} E_\mathrm{ADC} + f_s E_\mathrm{demod} + f_\mathrm{ds} E_\mathrm{MF} \Big) + E_\mathrm{NN}
\end{equation}
where $E_\mathrm{ADC}$ is the energy per ADC conversion, $E_\mathrm{demod}$ the demodulation and block-averaging energy per sample, $E_\mathrm{MF}$ the MF and RMF energy per I/Q block average, and $E_\mathrm{NN}$ the energy of the NN inference after the readout window. DIRAC spends
\begin{equation}\label{eq:edirac}
E_\mathrm{DIRAC} = t_\mathrm{meas} \big( P_\mathrm{ana} + 2 f_\mathrm{ds} E_\mathrm{TDC} + f_\mathrm{ds} E_\mathrm{MF} \big) + E_\mathrm{NN}
\end{equation}
where $P_\mathrm{ana}$ is the static power of the analog chain and $E_\mathrm{TDC}$ the energy per TDC conversion. Conversions run at $f_\mathrm{ds}$, 25$\times$ below $f_s$, where a counter costs far less than a Nyquist-rate ADC, so $E_\mathrm{TDC} \ll E_\mathrm{ADC}$. Demodulation and block averaging cost static power instead of energy per sample, and small bias currents meet the gain, dynamic range and bandwidth that the IF signal requires. Since $f_s$ drops out, the cost per qubit no longer grows with $M$. The analog chain and the TDCs draw only 44.2\,\textmu{W} per qubit while measuring and spend 61.2$\times$ less energy than the ADC, demodulation and blk-avg stages of FxP (\cref{sec:eval-energy}). The MF and the NN, only 15\% of the FxP RX BB energy, would then dominate; they stay digital to remain trainable, and \cref{sec:learning} cuts their cost with architecture--circuit co-designs.

DIRAC is ADC-less in that no Nyquist-rate ADC remains, but no function of an ADC is missing: the ping-pong integrators sample and hold, the V-to-Pulse stage compares each held value against a shared ramp, and the TDC digitizes, all at $f_\mathrm{ds}$. In more aggressive designs, the pulse interface could also support time-domain computation that replaces the TDCs and part of the digital logic. The following subsections describe each block.

\subsection{Analog Demodulation}\label{sec:arch-mixer}
For qubit $q$ with tone frequency $f_q$ and the complex input $x[n] = I[n] + jQ[n]$, the baseline's demodulator computes
\begin{equation}\label{eq:demod}
\begin{split}
y_q[n] &= x[n]\, e^{-j 2\pi f_q n / f_s} \\
       &= x[n] \big[\cos(2\pi f_q n / f_s) - j \sin(2\pi f_q n / f_s)\big]
\end{split}
\end{equation}

\begin{figure}[t]
    \centering
    \includegraphics[width=0.88\columnwidth]{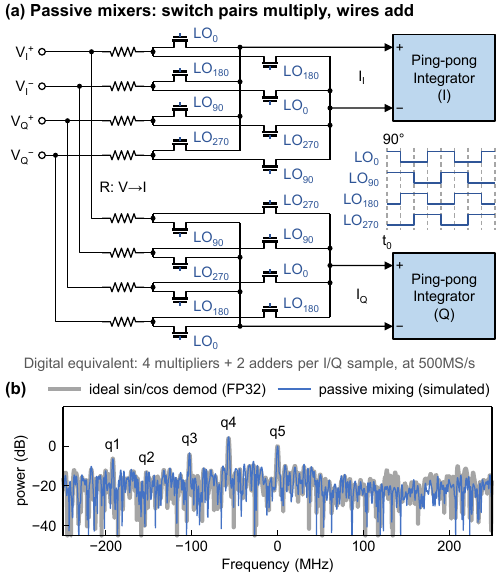}
    \Description{}
    \vspace{-4pt}
    \caption{Analog demodulation: (a) micro-architecture of the complex mixer, and (b) spectrum of the running example after square-wave mixing vs. ideal demodulation.}
    \label{fig:mixer}
\end{figure}

In DIRAC, a current-driven passive complex mixer (\cref{fig:mixer} (a)) takes over this multiplication. The IF buffers deliver I and Q as differential signals at $\sim$0.37\,V common mode, at most $\sim$170\,mV peak to peak on the dataset. The integrators hold the mixer outputs at a virtual ground with the same common mode, just as a transimpedance amplifier does behind a current-driven passive mixer~\cite{mirzaei2009analysis}, so the current through the resistors $R$ is just the AC swing of the buffer outputs divided by $R$. Four-phase LOs, square waves in phase with the cos/sin of \cref{eq:demod}, switch each current to the $+$ or $-$ output, i.e., multiply it by $\pm 1$, and the output wires add the products (at most $\sim$1.4\,\textmu{A} peak to peak on the dataset), so the mixer computes
\begin{equation}\label{eq:mix}
\begin{split}
i_q(t) &= I_I(t) + jI_Q(t) = \frac{x(t)\, \mathrm{LO}_q(t)}{R}, \\
\mathrm{LO}_q(t) &= \mathrm{sgn}\big(\cos(2\pi f_q t)\big) - j\,\mathrm{sgn}\big(\sin(2\pi f_q t)\big)
\end{split}
\end{equation}
where $x(t) = V_I(t) + jV_Q(t)$. The fundamental frequency component of $\mathrm{LO}_q(t)$ is $\tfrac{4}{\pi} e^{-j2\pi f_q t}$, so the mixer performs the demodulation of \cref{eq:demod} in the current domain and in continuous time. It is the equivalent of a complex digital multiplier, four multiplications and two additions per sample, but draws no power apart from its LO buffers ($<$0.1\,\textmu{W} per qubit). The square wave also has odd harmonics, with 1/3, 1/5, \ldots{} of the fundamental's amplitude at $-3f_q$, $5f_q$, \ldots, so the tones of a readout line must be planned with care. In the dataset used, only q3's harmonics come near other tones (q1 and q5), a fixed crosstalk. \cref{fig:mixer} (b) shows the mixer output of our running example, a trace that later figures reuse: mixing with q5's LO moves q5's tone to DC, and the spectrum matches ideal demodulation, with every tone's power within 1\,dB. The LOs come from the TX BB (\cref{sec:arch-lo}).

\subsection{Analog Block Averaging}\label{sec:arch-intg}
\begin{figure}[t]
    \centering
    \includegraphics[width=0.96\columnwidth]{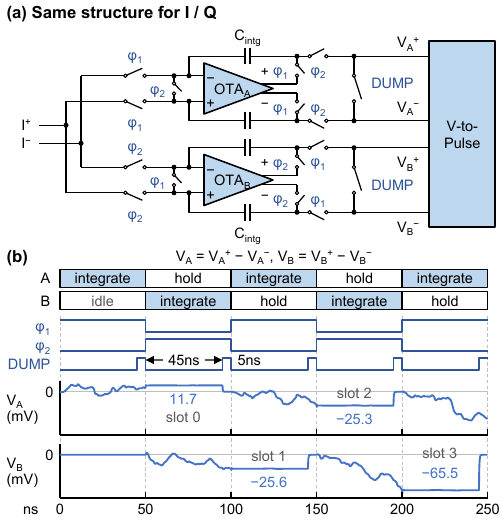}
    \Description{}
    \vspace{-4pt}
    \caption{Analog block averaging: (a) micro-architecture of a ping-pong integrator, and (b) its timing and operation on the I component of the running example.}
    \label{fig:intg}
\vspace{-8pt}
\end{figure}

After demodulation, the baseline's blk-avg stage computes
\begin{equation}\label{eq:blkavg}
s_q[k] = \frac{1}{N_\mathrm{ds}} \sum_{n=kN_\mathrm{ds}}^{(k+1)N_\mathrm{ds}-1} y_q[n]
\end{equation}
with $N_\mathrm{ds}$ = 25 samples per slot. DIRAC replaces it with an active-RC integrator: an operational transconductance amplifier (OTA) collects the mixer current on a capacitor $C_\mathrm{intg}$ during one slot and then holds the integrated voltage for conversion. To first order, this complex voltage is
\begin{equation}\label{eq:hold}
v_q[k] = \frac{\eta}{C_\mathrm{intg}} \int_{k T_\mathrm{slot}}^{(k+1)T_\mathrm{slot}} i_q(t)\,dt
\end{equation}
which computes the block average of \cref{eq:blkavg} in continuous time, up to a fixed gain. Here, $\eta \approx 0.67$ is the fraction of the mixer current that reaches $C_\mathrm{intg}$: the low-power OTA holds its input at a virtual ground only approximately, so the integrator's input impedance, in series with $R$, reduces the current by a fixed factor. This gain, the LO harmonics and device mismatch make the analog chain differ from the baseline, but they are static for a given chip, and DIRAC lets the discriminator learn them (\cref{sec:learning}).

The held value must stay constant while the V-to-Pulse stage converts it, and the integrator must then be reset, so a single integrator would miss half of the readout. Each ping-pong integrator therefore alternates two integrators (\cref{fig:intg}), so that, as in the baseline, no input is lost: while one integrates slot $k$, the other holds slot $k-1$ for conversion and is reset (DUMP) in the last 5\,ns of the slot. Over all training traces, the differential held values stay within $\sim\pm$190\,mV, i.e., each output moves by at most $\sim\pm$95\,mV around its 0.2\,V common mode, where the integrators remain nearly linear, and the four integrators of a qubit take 25.6\,\textmu{W} (6.4\,\textmu{W} each), the largest share of the analog chain.

\subsection{Time-Domain Digitization}\label{sec:arch-conv}
\begin{figure}[t]
    \centering
    \includegraphics[width=0.92\columnwidth]{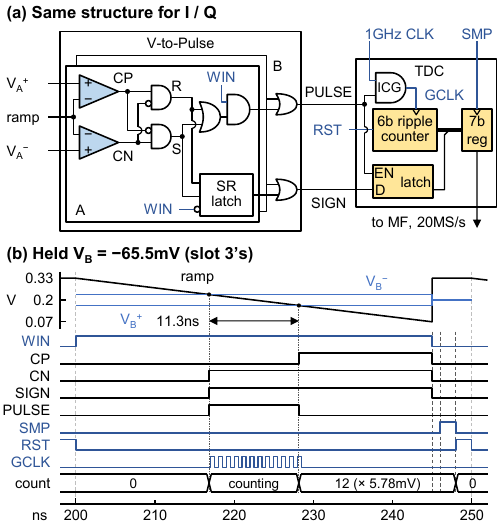}
    \Description{}
    \vspace{-4pt}
    \caption{Time-domain digitization: (a) micro-architecture of the V-to-Pulse stage and the TDC, and (b) waveforms of one conversion.}
    \label{fig:vtp}
\end{figure}

Each block average is held for a full 50\,ns slot, long enough to digitize it in the time domain rather than with an ADC (\cref{fig:vtp}). Each of I and Q has one V-to-Pulse stage, adapted from an analog floating-point FFT~\cite{chang2026afax}, which reads whichever of its two integrators holds and compares the two sides of the held differential value, e.g., $v_{q,I}[k] = \mathrm{Re}(v_q[k])$ ($V_A$/$V_B$ in \cref{fig:intg}), with a falling ramp of slope $S$ = 5.78\,mV/ns. The ramp is shared by all qubits of a line, and potentially by several lines; it falls from 0.33 to 0.07\,V to leave margin for common-mode shifts and comparator offsets on mismatched chips. Each comparator flips when the ramp crosses its input, so the time between the two flips is $|v_{q,I}[k]|/S$: PULSE is high in between, and SIGN records which comparator flipped first. The pulse width is then digitized by a simple ripple-counter TDC: it gates a clock of period $T_\mathrm{clk}$ with PULSE and counts its edges,
\begin{equation}\label{eq:count}
\tilde{s}_{q,I}[k] = \mathrm{SIGN} \cdot N_\mathrm{edge} \approx \frac{v_{q,I}[k]}{S\, T_\mathrm{clk}}, \qquad S\,T_\mathrm{clk} = 5.78\,\mathrm{mV}
\end{equation}
where $N_\mathrm{edge}$, the number of clock edges inside the pulse, differs from $|v_{q,I}[k]|/(S\,T_\mathrm{clk})$ by less than one. With Q converted likewise, $\tilde{s}_q[k] = \tilde{s}_{q,I}[k] + j\tilde{s}_{q,Q}[k]$ is, to first order, the block average $s_q[k]$ of \cref{eq:blkavg} times a fixed gain. We take 1\,GHz as a natural clock: the baseline already runs at 500\,MHz, and both systems can take a 1\,GHz clock from a PLL shared by many readout lines and other QCI blocks. The usable conversion window is about 41\,ns, the 45\,ns ramp minus the comparators' start-up and delay, so the TDC counts to about 40, and a 6b counter with the sign suffices (\cref{sec:eval-acc}). Because the ramp and the clock are shared, conversion is cheap: per qubit, the V-to-Pulse stages take 1.1\,\textmu{W} and the TDCs 2.1\,\textmu{W}.

\subsection{TX--RX Baseband Co-Design}\label{sec:arch-lo}
\begin{figure}[t]
    \centering
    \includegraphics[width=0.92\columnwidth]{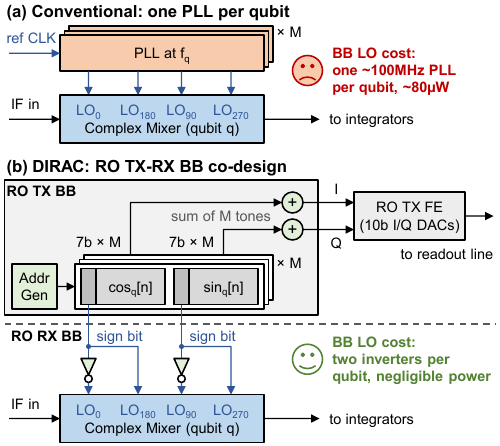}
    \Description{}
    \vspace{-4pt}
    \caption{BB LO generation for analog demodulation: (a) conventional, using one PLL per qubit, and (b) DIRAC, from the NCO sign bits of readout TX BB.}
    \label{fig:lo}
\end{figure}

Analog demodulation needs four phases of a square-wave LO at each qubit's tone frequency. A stand-alone RX would synthesize them with one PLL per tone, each costing about 80\,\textmu{W} at BB frequencies~\cite{huang2025cryogenic} (\cref{fig:lo} (a)). We instead envision the readout TX BB and RX BB of a line on one chip, which also reduces the system footprint, and take the LOs from the TX BB (\cref{fig:lo} (b)). However it is designed, a readout TX BB has a numerically controlled oscillator (NCO) per tone, e.g., a phase accumulator with cos/sin LUTs~\cite{min2023qisim}, CORDIC units, or tables of stored cos/sin words oscillating at exactly the frequency RX needs. The square-wave LOs of \cref{eq:mix} are the signs of these words. With $b_c[n]$ and $b_s[n]$ the sign bits (two's-complement MSBs) of tone $q$'s words $\cos_q[n] = \cos(2\pi f_q n T_\mathrm{clk})$ and $\sin_q[n] = \sin(2\pi f_q n T_\mathrm{clk})$,
\begin{equation}\label{eq:lo}
\begin{split}
\mathrm{LO}_{0}[n] &= \overline{b_c[n]}, \quad \mathrm{LO}_{180}[n] = b_c[n], \\
\mathrm{LO}_{90}[n] &= \overline{b_s[n]}, \quad \mathrm{LO}_{270}[n] = b_s[n]
\end{split}
\end{equation}
i.e., LO$_0$ is high when $\cos_q[n] \ge 0$ and LO$_{90}$ when $\sin_q[n] \ge 0$. These four phases are exactly the square waves of \cref{eq:mix}, so the RX BB LOs cost two inverters per qubit. Without loss of generality, our TX BB stores the cos/sin words of each tone in tables, replays them at 1\,GS/s, and sums the $M$ tones into the words of the I/Q DACs that the line's qubits share; at every cycle, the sign bits of the words it reads go to the RX BB mixers. Since the TX and the RX use the same words, read from the same address at every readout, the LOs are synchronized with the transmitted tones by construction, and every shot sees identical LO edges. Their edges fall on the 1\,ns grid of the TX BB (instead of ideal continuous time), a fixed non-ideality that the discriminator learns.

\section{Learned Calibration and Co-Design}\label{sec:learning}
\begin{figure}[t]
    \centering
    \includegraphics[width=0.84\columnwidth]{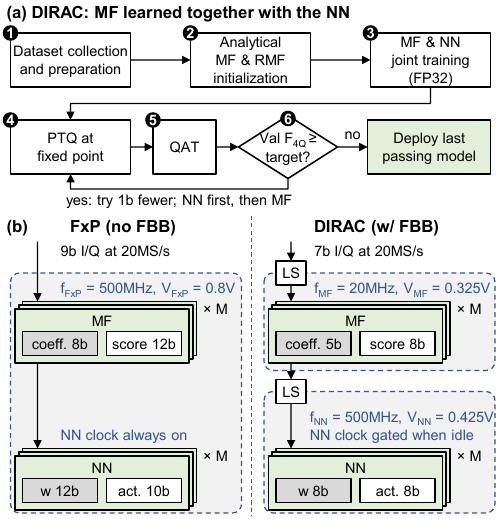}
    \Description{}
    \vspace{-4pt}
    \caption{MF and NN: (a) the per-line calibration flow with trained MF, and (b) hardware designs of FxP vs.\ DIRAC.}
    \label{fig:digital}
\end{figure}

\subsection{Per-Line Calibration with Trained MF}\label{sec:learn-mf}\label{sec:learn-quant}
Analog computing is imprecise. To first order, the analog chain computes the baseline's block averages times a fixed gain (\cref{sec:arch-intg}), but its gain differs between qubits and between I and Q through within-die variation, the switching of its LOs and integrators adds offsets and crosstalk between slots, and the V-to-Pulse stage and the TDC add a slightly nonlinear transfer curve; die-to-die variation makes all of them different on every chip. Trimming each of them with extra circuits in the analog chain would cost area, power and test time. DIRAC instead lets the discriminator learn them: they stay the same in every shot of a given line and its fabricated readout hardware, so a discriminator trained on the line's own outputs $\tilde{s}_q[k]$ sees them as fixed features of the data.

\cref{fig:digital} (a) shows the calibration flow, which runs offline once per readout line on labelled shots of prepared states; readout calibration and discriminator training need these with or without an analog chain (e.g., the training shots of the dataset~\cite{lienhard2022deep}). For each qubit, the MF and the RMF are first initialized analytically from the line's own TDC outputs, as in the baseline~\cite{lienhard2022deep,maurya2023herqules}, and then trained jointly with the NN heads in FP32: the MF becomes a learned first layer, matched to the chain that produces its inputs. Training recovers +0.41\% of $F_{4Q}$ over the closed-form MF (\cref{sec:eval-acc}) and is repeated whenever the discriminator is recalibrated anyway, e.g., when qubits or readout signals drift~\cite{google2025quantum}.

The same flow allows aggressive quantization. Starting from the trained FP32 model, it first applies post-training quantization to high-bit FxP that loses no accuracy, e.g., near the FxP baseline's widths, and then removes one bit at a time, first from the NN, which holds most of the weights and compute, then from the MF; after each step, it fine-tunes the model with QAT and checks the validation $F_{4Q}$ against the FP32 digital baseline's, and the last model that passes is deployed. On the worst of our ten mismatch chips, the flow ends at 5b MF coefficients and 8b NN weights, against 8b and 12b in the FxP baseline (\cref{sec:method-acc,sec:eval-acc}).

\subsection{Architecture--Circuit Co-Designs}\label{sec:digital}
What remains digital is small, and its low bit widths also shorten its critical paths; DIRAC exploits this with three techniques across architecture and circuit (\cref{fig:digital} (b)). (1) \emph{Multi-clock, multi-supply design}: the MF runs at the rate of the block averages, $f_\mathrm{ds}$ = 20\,MHz, and the NN heads still at 500\,MHz to keep the baselines' readout latency, each in its own clock and supply domain, joined by level shifters. (2) \emph{Architectural clock gating}: the NN heads are clocked only during their 60\,ns burst at the end of each readout, an idleness that the fine-grained clock gating of synthesis cannot see. (3) \emph{Low-voltage operation with FBB}: each domain runs at the lowest supply at which its critical paths still meet timing, lowered further by 2\,V of FBB: 0.325\,V for the MF and 0.425\,V for the NN heads (\cref{sec:method-hw}). Together, these techniques bring the MF and the NN to 56.4\,\textmu{W} per qubit against 471\,\textmu{W} in the FxP baseline (\cref{sec:eval-energy}).

\section{Evaluation Methodology}\label{sec:method}

\subsection{Hardware Characterization}\label{sec:method-hw}
All systems assume $M$ = 6 qubits per readout line, as in Google's quantum processor~\cite{google2023suppressing}. The baselines' ADCs are modeled by the 10b, 500\,MS/s mode of a measured 4\,K ADC~\cite{lee2023programmable}, 1.97\,mW per qubit while measuring. We design the remaining RX BB blocks and one readout TX BB for the evaluated cryo-CMOS systems in 22\,nm FD-SOI with a commercial 4\,K device model. Digital designs are synthesized, placed and routed with a standard-cell library re-characterized at 4\,K and 0.8\,V, with wire resistance scaled to 0.7$\times$ its RT value~\cite{saligram2021scaled} for 4\,K, and their power is simulated post-layout on 256 test readouts.

We design and size DIRAC's analog chain for one qubit as a layout-ready schematic for the dataset's traces, up-sampled from 500\,MS/s to emulate the RX FE's continuous-time output: every stage keeps headroom for the largest swing in the training set. HSPICE simulation at 4\,K gives the power of the chain (at 0.8\,V) and its TDCs (at 0.6\,V). DIRAC's digital logic (MF, NN and TX BB, which the baselines run at 0.8\,V) runs at lower supplies: since the standard-cell library exists only at 0.8\,V, we simulate the post-layout critical paths of each domain in HSPICE with FBB, take the lowest supply $V$ that meets its clock period, and scale the domain's 0.8\,V power by $(V/0.8\,\mathrm{V})^2$, as leakage is negligible at 4\,K. We exclude the global readout controller, global clock generation (500\,MHz for the baselines, 1\,GHz for DIRAC), and DIRAC's ramp and bias generators, all shared by many readout lines.

\subsection{Readout Fidelity}\label{sec:method-acc}
\textbf{Dataset and discriminator.} We use the first 1\,\textmu{s} of the dataset in~\cite{lienhard2022deep}: 50\,k shots for each of 32 labels (five qubits) on one line. As we do not discriminate q2, labels that differ only in its state are merged, giving 16 labels of 100\,k shots each. Per label, 20\,k shots are for training, 10\,k for validation and 70\,k for testing, 320\,k, 160\,k and 1.12\,M in total. Model selection, e.g., of bit widths, uses only validation fidelity. Every design uses the discriminator of \cref{sec:mot-pipeline}, trained from scratch for every design and readout window length. The baselines keep the closed-form MF, as in~\cite{lienhard2022deep,maurya2023herqules,mude2025efficient}.

\begin{table}[t]
\caption{Behavioral model of DIRAC's analog chain vs.\ circuit simulation: mean error per slot value, in TDC LSB.} \vspace{-8pt}
\label{tab:modelacc}
\centering\footnotesize
\setlength{\tabcolsep}{4pt}
\begin{tabular}{|l|r|r|r|r|r|}
\hline
RX BB LO & q1 & q3 & q4 & q5 & Mean \\
\hline
Ideal LO\textsuperscript{a} & 0.047 & 0.047 & 0.045 & 0.045 & 0.046 \\
Co-designed LO\textsuperscript{b} & 0.050 & 0.057 & 0.057 & 0.053 & 0.054 \\
\hline
\multicolumn{6}{@{}l}{\footnotesize\textsuperscript{a}\,Nominal chain. \textsuperscript{b}\,Mean of all 11 chains.} \\
\end{tabular}
\vspace{-4pt}
\end{table}

\textbf{Behavioral model of the analog chain.} Transistor-level simulation of one 1\,\textmu{s} trace through one qubit's analog chain takes about 7 minutes, so simulating the dataset's 1.6\,M shots is intractable and needs a behavioral model. We build it from the function of each circuit block and fit its parameters to dedicated circuit simulations, never to the traces it is validated on. Noise is modeled as carefully: 4\,K noise simulations of the chain's devices, e.g., the thermal noise of its resistors and switches, give the noise statistics of every slot, from which the model draws a random sample in every slot of every shot. The model takes about 20\,\textmu{s} per trace and qubit, seven orders of magnitude faster than transistor-level simulation.

Besides the nominal chain, we draw ten mismatched chains from the local mismatch of the 4\,K device model and model each separately, as every fabricated chip differs; the nominal chain is modeled with both the ideal square-wave LO and the co-designed LO, the others with the latter. On 40 traces per qubit not used to build the models, i.e., 6,400 slot values per chain over the four qubits, they follow transistor-level simulation to about 0.05 LSB of the TDC output (\cref{tab:modelacc}; 1 LSB = 5.78\,mV), under 2\% of the about 3 LSB of noise per slot in the dataset's own traces.

\textbf{Composite chips.} Assuming one chip per readout line, a \emph{composite chip} takes four different mismatched chains, one per qubit, as on a real die. We form ten composite chips from different sets of chains and run each through the calibration flow of \cref{sec:learning}, with the dataset's traces passed through the chip's model, noise included. For DIRAC, we report the mean and standard deviation of $F_{4Q}$ over the ten composite chips.

\subsection{4\texorpdfstring{\,}{ }K QCI System Power Modeling}\label{sec:method-system}

\begin{table}[t]
\caption{4\,K power of the blocks not designed in this work.} \vspace{-8pt}
\label{tab:sysinputs}
\centering\footnotesize
\setlength{\tabcolsep}{3pt}
\begin{tabular}{|L{0.2\columnwidth}|L{0.6\columnwidth}|L{0.08\columnwidth}|}
\hline
RT QCI & Power & Ref. \\
\hline
Coax cable & 1.0\,mW per XY or RO cable, 1.2\,mW per flux cable & \cite{krinner2019engineering} \\
Flex cable & 0.315\,mW per cable & \cite{min2023qisim} \\
Attenuators & 7.9\,\textmu{W} per cable, counted as always-on & \cite{min2023qisim} \\
HEMT & 7.8\,mW per RO line & \cite{shresthamali2026revisiting} \\
\hline
Cryo-CMOS QCI & Power & Ref. \\
\hline
XY / Z control & $P_\mathrm{XY}$ = 2.4\,mW/qubit, $P_\mathrm{Z}$ = 0.9\,mW/qubit or coupler & \cite{yoo2023design} \\
RO TX FE & 1.83\,mW per RO line, 0.30\,mW/qubit & \cite{chakraborty2024rfawg,vanwinckel2022cryo} \\
RO RX FE & 0.34\,mW/qubit & \cite{nagulu2023sub} \\
ADC (baselines) & 2 $\times$ 5.91\,mW per RO line, 1.97\,mW/qubit & \cite{lee2023programmable} \\
\hline
\end{tabular}
\end{table}

We estimate the 4\,K power of the complete QCI of an LQ patch per ESM round. For an RT QCI, it is the static heat of the cables thermally anchored at 4\,K and their attenuators, plus one HEMT per readout line (\cref{tab:sysinputs}). We follow the wiring of Google's processor~\cite{yoo2023design,google2023suppressing}: every qubit has its own XY and Z cable and every coupler its own flux cable, two per qubit on average; with an input and an output cable per readout line of $M$ = 6 qubits, this is 4.3 cables per qubit. We favor RT QCIs by ignoring the HEMT bias wiring.

The cryo-CMOS blocks we do not design are taken from measured 4\,K chips (\cref{tab:sysinputs}): the XY and Z controllers~\cite{yoo2023design}; the readout TX FE, I/Q DACs~\cite{chakraborty2024rfawg} plus the RF part of a readout TX, estimated from~\cite{vanwinckel2022cryo}; the RX FE~\cite{nagulu2023sub}; and the baselines' ADC~\cite{lee2023programmable}. Shared RF LOs are excluded, as in the XY and Z controller. The RX FE replaces the HEMT; the few digital links to RT, shared by many LQs, are not counted. As in QIsim~\cite{min2023qisim}, each block contributes its power times its active time in a round of Google's schedule (\cref{sec:background-qec}), averaged over the round; control and readout draw power only while active. For a distance-$d$ patch, a cryo-CMOS QCI draws
\begin{equation}\label{eq:patch}
\begin{split}
P_\mathrm{patch}(d) = \frac{1}{T_\mathrm{rnd}} \Big[ & 5(2d^2-1) \cdot t_\mathrm{1q} P_\mathrm{XY}
+ 4d(d-1) \cdot t_\mathrm{cz} \cdot 3P_\mathrm{Z} \\
& + (d^2-1) \cdot \big(t_\mathrm{rst} P_\mathrm{Z} + t_\mathrm{meas} P_\mathrm{RO}\big)\Big]
\end{split}
\end{equation}
where each of the $2d^2-1$ qubits receives five single-qubit gates per round, the patch has $4d(d-1)$ CZ gates, each pulsing three Z controllers (qubits and coupler)~\cite{yoo2023design,google2023suppressing}, and $P_\mathrm{RO}$ is the per-qubit power of the readout TX and RX together while measuring; \cref{sec:eval-qec} uses \cref{eq:patch} for the exact gain in code distance. Dividing \cref{eq:patch} by the $2d^2-1$ qubits gives the power per physical qubit. As $d$ grows, it approaches its value in the bulk of a large patch, $(P_\mathrm{DQ}+P_\mathrm{AQ})/2$, where DQs and AQs are 1:1 and each is in four CZ gates. A finite patch has fewer CZ gates per qubit, as qubits on its boundary are in fewer of them, so its qubits draw less on average: the bulk value is an upper bound. \cref{sec:eval-system} uses it to estimate the gain in system scale, i.e., how many more qubits a 4\,K budget holds.

\begin{table}[t]
\caption{Readout fidelity at 1\,\textmu{s} window length.} \vspace{-8pt}
\label{tab:accuracy}
\centering\footnotesize
\setlength{\tabcolsep}{2.4pt}
\begin{tabular}{|L{0.36\columnwidth}|r|r|r|r|c|r|}
\hline
Design & q1 & q3 & q4 & q5 & $F_{4Q}$ & $\Delta$ \\
\hline
FP32 baseline & 97.00 & 92.73 & 94.49 & 96.99 & 95.28 & -- \\
FxP baseline & 97.02 & 92.69 & 94.50 & 96.98 & 95.28 & 0.00 \\
\hline
Nom. chip, ideal LO\textsuperscript{a} & 96.77 & 92.26 & 94.41 & 96.95 & 95.08 & $-$0.20 \\
Nom. chip, co-designed LO\textsuperscript{a} & 96.70 & 92.31 & 94.38 & 96.92 & 95.06 & $-$0.22 \\
\hline
Comp. chips, co-designed LO\textsuperscript{a} & 96.69 & 92.25 & 94.39 & 96.92 & 95.04$\pm$0.01\textsuperscript{b} & $-$0.24 \\
\quad + trained MF\textsuperscript{a} & 96.67 & 93.72 & 94.54 & 96.90 & 95.45$\pm$0.01\textsuperscript{b} & +0.17 \\
\quad + MF 5b, NN 8b (DIRAC) & 96.61 & 93.60 & 94.47 & 96.87 & \textbf{95.38$\pm$0.02}\textsuperscript{b} & +0.10 \\
\hline
\multicolumn{7}{@{}l}{\footnotesize\textsuperscript{a}\,FP32 arithmetic; closed-form MF unless noted.} \\
\multicolumn{7}{@{}l}{\footnotesize\textsuperscript{b}\,Mean $\pm$ std.\ over ten composite chips.} \\
\end{tabular}
\end{table}

\subsection{Logical Error Rate}\label{sec:method-ler}
To evaluate how the larger system scale of DIRAC improves QEC performance, we simulate a rotated surface-code memory in the Z basis with Stim~\cite{gidney2021stim}, $d$ rounds per shot and CZ-compiled circuits as on Google's hardware, under two noise models: \emph{Willow-72}, the mean component errors measured on Google's 72-qubit processor~\cite{google2025quantum}, i.e., today's hardware, decoded with correlated matching~\cite{fowler2013optimal} as in Google's analysis, and \emph{SI1000} at $p = 10^{-3}$ for near-term hardware~\cite{gidney2022benchmarking}, decoded by minimum-weight perfect matching (MWPM)~\cite{higgott2025sparse}. The baseline is the unmodified noise model, i.e., we assume that both RT QCI and FxP reach the readout fidelity of the noise models. DIRAC's measurement error is the noise model's error scaled by the ratio of its readout infidelity to the FxP baseline's at 500\,ns (\cref{sec:eval-acc}), with all other errors following the noise models. We report the LER per logical cycle, i.e., per $d$-round shot decoded once.
Beyond the simulated distances, the baseline's LER per round is extrapolated with the standard fit $\varepsilon_d = A\Lambda^{-(d+1)/2}$ to its simulated values~\cite{google2025quantum}, and DIRAC's is the baseline's times DIRAC's mean increase over the distances simulated for both; both are then converted to the LER per cycle ($d$ rounds).

\section{Evaluation Results}\label{sec:eval}

\begin{table*}[t]
\caption{4\,K cryo-CMOS QCI power per physical qubit (bulk) over one ESM round.} \vspace{-8pt}
\label{tab:power4k}
\centering\small
\setlength{\tabcolsep}{8pt}
\begin{tabular}{|c|l|l|r|r|r|r|r|r|r|}
\hline
\multirow{2}{*}{Qubit} & \multirow{2}{*}{Op.} & \multirow{2}{*}{Block} & \multirow{2}{*}{$t$ (ns)} & \multicolumn{3}{c|}{Active power (mW)} & \multicolumn{3}{c|}{Energy per round (pJ)} \\
\cline{5-10}
 & & & & FP32 & FxP & DIRAC & FP32 & FxP & DIRAC \\
\hline
\multirow{3}{*}{DQ} & 5$\times$1Q & XY controller & 125 & \multicolumn{3}{c|}{2.40} & \multicolumn{3}{c|}{300} \\
\cline{2-10}
 & 4$\times$CZ & Z controller & 136 & \multicolumn{3}{c|}{1.35} & \multicolumn{3}{c|}{184} \\
\cline{2-10}
 & \multicolumn{3}{l|}{\textbf{DQ total\textsuperscript{a} ($P_\mathrm{DQ}$)}} & \multicolumn{3}{c|}{\textit{0.53}} & \multicolumn{3}{c|}{484} \\
\hline
\multirow{8}{*}{AQ} & 5$\times$1Q & XY controller & 125 & \multicolumn{3}{c|}{2.40} & \multicolumn{3}{c|}{300} \\
\cline{2-10}
 & 4$\times$CZ & Z controller & 136 & \multicolumn{3}{c|}{1.35} & \multicolumn{3}{c|}{184} \\
\cline{2-10}
 & \multirow{4}{*}{Measure} & RO TX BB & 500 & 0.25 & 0.25 & 0.16 & 127 & 127 & 80 \\
\cline{3-10}
 &  & RO TX FE & 500 & \multicolumn{3}{c|}{0.30} & \multicolumn{3}{c|}{152} \\
\cline{3-10}
 &  & RO RX FE & 500 & \multicolumn{3}{c|}{0.34} & \multicolumn{3}{c|}{170} \\
\cline{3-10}
 &  & RO RX BB & 500 & 23.96 & 3.18 & 0.10 & 11{,}981 & 1{,}589 & 50 \\
\cline{2-10}
 & Reset & Z controller & 160 & \multicolumn{3}{c|}{0.90} & \multicolumn{3}{c|}{144} \\
\cline{2-10}
 & \multicolumn{3}{l|}{\textbf{AQ total\textsuperscript{a} ($P_\mathrm{AQ}$)}} & \textit{14.18} & \textit{2.89} & \textit{1.17} & 13{,}058 & 2{,}666 & 1{,}080 \\
\hline
\multicolumn{4}{|l|}{\textbf{Per physical qubit}\textsuperscript{a}, $(P_\mathrm{DQ} + P_\mathrm{AQ})/2$} & \textbf{\textit{7.35}} & \textbf{\textit{1.71}} & \textbf{\textit{0.85}} & 6{,}771 & 1{,}575 & 782 \\
\hline
\multicolumn{10}{@{}l}{\footnotesize\textsuperscript{a}\,Average power over the 921\,ns ESM round.} \\
\end{tabular}
\end{table*}

\subsection{Readout Fidelity}\label{sec:eval-acc}
\begin{figure}[t]
    \centering
    \includegraphics[width=0.96\columnwidth]{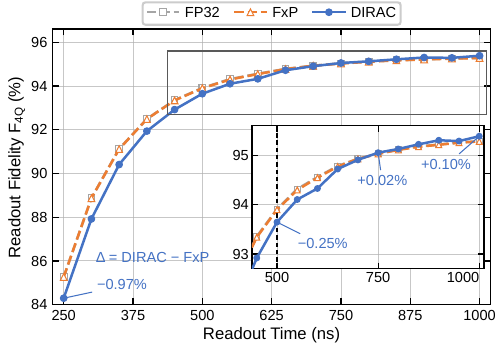}
    \Description{}
    \vspace{-4pt}
    \caption{Readout fidelity vs.\ time, with every design retrained at every window length.}
    \label{fig:rotime}
\end{figure}

\cref{tab:accuracy} gives the readout fidelity of every design on the 1.12\,M test shots at the dataset's 1\,\textmu{s} readout. FxP keeps the fidelity of FP32 (95.28\%). On the nominal chip with an ideal LO, the analog chain changes $F_{4Q}$ by $-$0.20\%, mainly through the square-wave LO's harmonics and the chain's static errors; the co-designed LO changes it by a further $-$0.02\% through its static phase errors, and mismatch across the ten composite chips by $-$0.02\%. Training the MF recovers +0.41\%, mostly on q3, which loses the most in the analog chain. Quantizing the MF to 5b and the NN to 8b then changes $F_{4Q}$ by $-$0.07\%, which leaves DIRAC at 95.38\,$\pm$\,0.02\%, above FP32; even its worst composite chip reaches 95.34\%. This margin does not allow narrower widths: they are chosen on validation shots (\cref{sec:learn-quant}), where the worst composite chip stays just above FP32 at these widths and falls below it with a 4b MF.

\begin{figure}[t]
    \centering
    \includegraphics[width=0.8\columnwidth]{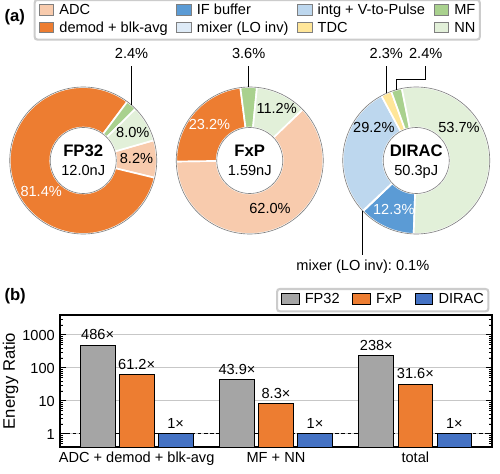}
    \Description{}
    \vspace{-4pt}
    \caption{Readout RX BB energy per readout (500\,ns, $M$ = 6): (a) breakdown, and (b) the baselines relative to DIRAC.}
    \label{fig:energy}
\vspace{-8pt}
\end{figure}

\cref{fig:rotime} sweeps the readout window from 250 to 1000\,ns, with DIRAC on the composite chip closest to the ten-chip mean. DIRAC matches the FP32 fidelity down to 750\,ns. At 500\,ns, the readout time of Google's processor (\cref{sec:background-qec}), it is 0.25\% below FxP (93.65\% vs.\ 93.90\%), a gap that \cref{sec:eval-qec} carries into the LER. The gap widens at shorter windows, mostly due to the square-wave LO, whose harmonics also demodulate noise.

\subsection{Readout RX Baseband Hardware Overheads}\label{sec:eval-energy}
Following Google's setup, \cref{fig:energy} (a) breaks down the RX BB energy of one 500\,ns readout for $M$ = 6, calculated by multiplying every block's power by the readout time and adding the NN's 60\,ns burst energy, with DIRAC's analog chain and TDCs running one slot longer (the last slot's digitization); the power of the NN, built for the dataset's four qubits (8-8-8-2), is scaled by the MAC ratio of a 12-8-8-2 head (1.2$\times$). DIRAC takes 50.3\,pJ per qubit, 31.6$\times$ and 238$\times$ less than FxP (1.59\,nJ) and FP32 (12.0\,nJ); while measuring, i.e., averaged over the readout time, it draws 100.6\,\textmu{W} per qubit, against 3.18\,mW and 24.0\,mW. The NN and the integrators dominate, while the passive mixer draws almost nothing.

\cref{fig:energy} (b) separates the two sources of this reduction. Up to the TDCs, DIRAC needs 61.2$\times$ less energy than the ADCs, demodulation and block averaging of FxP (486$\times$ less than those of FP32), the effect of the redrawn A/D boundary. Its MF and NN need 8.3$\times$ (43.9$\times$) less, the effect of the low bit widths that per-line calibration permits and of the co-designs. The RX BB finishes 60\,ns after the readout window in the baselines and 110\,ns in DIRAC, within the following 160\,ns reset, so discrimination adds no backlog.

\subsection{4\texorpdfstring{\,}{ }K QCI System Power}\label{sec:eval-system}
\begin{figure}[t]
    \centering
    \includegraphics[width=0.9\columnwidth]{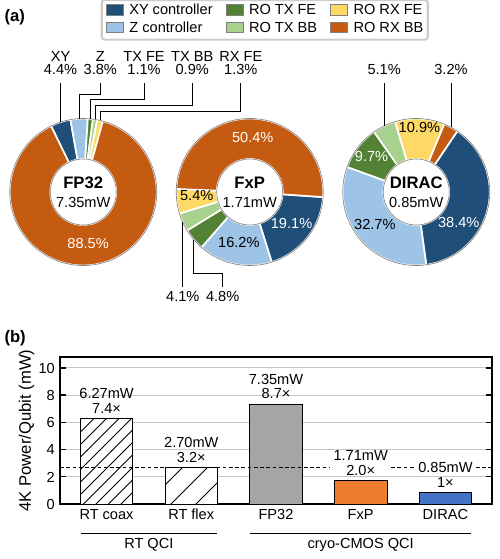}
    \Description{}
    \vspace{-4pt}
    \caption{4\,K QCI power per physical qubit: (a) breakdown for cryo-CMOS designs, and (b) comparison with RT QCIs.}
    \label{fig:system}
\end{figure}

To compute the bulk value of the cryo-CMOS QCI power per physical qubit, \cref{tab:power4k} applies the model of \cref{sec:method-system} to a bulk DQ and AQ over one ESM round; the Z power of a coupler is split between its two qubits, so each draws $1.5\,P_\mathrm{Z}$ during a CZ. The cryo-CMOS QCI with DIRAC draws 0.85\,mW per physical qubit, against 1.71\,mW with FxP and 7.35\,mW with FP32, a reduction of 50.4\% and 88.5\%. Any 4\,K budget thus holds 2.0$\times$ (8.7$\times$) as many physical qubits, and as many LQs at the same distance, a large gain in system scale wherever 4\,K power is the bottleneck. These bulk ratios are representative: for single patches of $d$ = 3 to 27, the exact ratios stay within 3.2\% of them. The three designs share the same controllers and readout FEs; the difference comes from the RX BB and, to a lesser extent, the TX BB, which DIRAC runs at 0.5\,V with FBB against 0.8\,V in the baselines (\cref{sec:method-hw}).

\cref{fig:system} (a) shows that the bottleneck moves. The RX BB takes 88.5\% (FP32) and 50.4\% (FxP) of the 4\,K power but only 3.2\% with DIRAC, where the qubit controllers dominate with 71.1\%. \cref{fig:system} (b) compares the cryo-CMOS designs with the 4\,K load of RT QCIs, i.e., the heat of their cables and HEMTs (\cref{sec:method-system}): DIRAC draws 3.2$\times$ less 4\,K power than using flex cables and 7.4$\times$ less than using coaxial cables. Cryo-CMOS readout thus pays off only when its power is optimized: FP32 draws more 4\,K power than either kind of cable, whereas DIRAC draws the least of all designs.

\subsection{Code Distance and Logical Error Rate}\label{sec:eval-qec}
\begin{figure}[t]
    \centering
    \includegraphics[width=1.0\columnwidth]{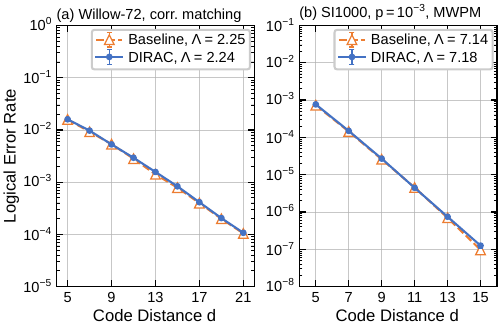}
    \Description{}
    \vspace{-4pt}
    \caption{LER vs.\ code distance.}
    \label{fig:ler}
\end{figure}

\cref{fig:ler} shows the LER of DIRAC and of the baseline (RT QCI or FxP), whose measurement error is that of each noise model, across $d$. The baseline's suppression factor $\Lambda$, the ratio of the LERs per round at distances $d$ and $d+2$, is 2.25 on Willow-72, within about 15\% of Google's measured values, so our simulation is realistic. At 500\,ns, DIRAC's readout infidelity is 1.041$\times$ the baseline's (FxP), which barely moves the LER at equal $d$: $\Lambda$ is 2.24 against 2.25 on Willow-72 and 7.18 against 7.14 on SI1000. Averaged over the $d$ simulated for both designs, DIRAC's LER at the same $d$ is 3.7\% (Willow-72) and 4.6\% (SI1000) higher, with no trend in $d$.

\begin{table}[t]
\caption{Distance $d'$ that DIRAC reaches at the 4\,K QCI power per LQ of a baseline at $d_0$, and the factor of LER reduction.} \vspace{-5pt}
\label{tab:dmap}
\centering\normalsize
\setlength{\tabcolsep}{7pt}
\begin{tabular}{|c|c|r|r|c|r|r|}
\hline
\multirow{3}{*}{$d_0$} & \multicolumn{3}{c|}{vs.\ RT flex} & \multicolumn{3}{c|}{vs.\ FxP} \\
\cline{2-7}
 & \multirow{2}{*}{$d'$} & \multicolumn{2}{c|}{LER reduction} & \multirow{2}{*}{$d'$} & \multicolumn{2}{c|}{LER reduction} \\
\cline{3-4}\cline{6-7}
 & & Willow & SI1000 & & Willow & SI1000 \\
\hline
5 & 7 & 1.6$\times$ & 4.9$\times$ & 7 & 1.6$\times$ & 4.9$\times$ \\
7 & 11 & 3.1$\times$ & 32$\times$ & 9 & 1.7$\times$ & 5.3$\times$ \\
9 & 15 & 6.4$\times$ & 206$\times$ & 11 & 1.8$\times$ & 5.8$\times$ \\
11 & 19 & 14$\times$ & 1{,}476$\times$\textsuperscript{a} & 15 & 3.4$\times$ & 36$\times$ \\
13 & 23 & 29$\times$\textsuperscript{a} & 9{,}604$\times$\textsuperscript{a} & 17 & 3.4$\times$ & 40$\times$\textsuperscript{a} \\
15 & 25 & 33$\times$\textsuperscript{a} & 8{,}730$\times$\textsuperscript{a} & 21 & 7.3$\times$ & 204$\times$\textsuperscript{a} \\
\hline
\multicolumn{7}{@{}l}{\footnotesize\textsuperscript{a}\,Extrapolated using the methodology of \cref{sec:method-ler}.} \\
\end{tabular}
\end{table}

When the 4\,K power budget limits the system scale, the 4\,K QCI power saving translates into a larger code distance at a fixed number of LQs. \cref{tab:dmap} gives, for a baseline patch of distance $d_0$, with RT QCI on flex cables or a cryo-CMOS QCI with FxP, the largest distance $d'$ at which a DIRAC patch draws no more 4\,K power (\cref{eq:patch}); entries beyond the simulated distances are extrapolated. The larger distance more than offsets DIRAC's slightly higher readout error. Against flex cables, $d'$ exceeds $d_0$ by 2 to 10, and DIRAC's LER per cycle is 1.6--33$\times$ lower on Willow-72 and 4.9--9{,}604$\times$ lower on SI1000; from $d_0$ = 11, it reaches $d'$ = 19 with a 14$\times$ and 1{,}476$\times$ lower LER per cycle. Against FxP, $d'$ exceeds $d_0$ by 2 to 6, and the LER per cycle is 1.6--7.3$\times$ and 4.9--204$\times$ lower; from $d_0$ = 11, DIRAC reaches $d'$ = 15 with a 3.4$\times$ and 36$\times$ lower LER per cycle.

\section{Related Work}\label{sec:related}

\textbf{Discrimination algorithms and hardware.} Discriminators range from MFs~\cite{lienhard2022deep}, trainable temporal filters~\cite{khan2024practical} and reservoir computing~\cite{kent2026superconducting} to NNs: fully connected networks on raw traces~\cite{lienhard2022deep}, MFs with small NNs~\cite{maurya2023herqules,mude2025efficient}, CNNs and transformers~\cite{giortamis2026mcmit}, recurrent~\cite{le2026realtime} and state-space models~\cite{otting2026multistage}, and compressed or quantized NNs~\cite{guo2025klinq,gautam2026quantized,diguglielmo2025endtoend}, many compared in Oraqle~\cite{giortamis2026oraqle}. Their hardware runs at RT behind the conventional ADC and DSP chain, mostly on FPGAs~\cite{guo2025klinq,gautam2026quantized,diguglielmo2025endtoend,giortamis2026mcmit,le2026realtime}; ASIC designs simulated at RT draw 1138\,mW~\cite{liu2025qubit} and 148\,mW after synthesis~\cite{du2026niaq}, far beyond the desired 4\,K target ($<$1\,mW per qubit).

\textbf{Cryogenic readout chips.} Measured cryogenic readout chips mostly implement the FE: RXs~\cite{nagulu2023sub,vanwinckel2022cryo,prabowo2021cryo,kwende2025design} and ADCs~\cite{kiene2021cryo,lee2023programmable,veraverbeke2025cryo} deliver analog or digitized I/Q. Chips that also discriminate either serve one qubit, like a reflectometry IC that does not support FDM by design~\cite{guo2026reflectometry}, or rule out ML discrimination with fixed signal processing, from brute-force integration over the window~\cite{minn2023cryo,minn2025cryo} to per-tone phase detection or thresholding~\cite{guo2024cryo,guo2026cryogenic,huang2025cryogenic,park2021fully}. No cryogenic qubit readout chip combines FDM with ML discrimination.

\textbf{Analog computing.} Like RedEye, which runs CNN layers before quantization~\cite{likamwa2016redeye}, and in-memory accelerators that cut their A/D cost with data encoding~\cite{shafiee2016isaac}, time-domain interfaces~\cite{li2020timely} or low-resolution conversion~\cite{andrulis2023raella}, DIRAC computes before the converter. DIRAC is also trained with its hardware's errors, but whereas hardware-aware training fits weights stored in analog devices to them~\cite{rasch2023hardware}, DIRAC redraws the A/D boundary so that every trained parameter stays digital: its fixed analog chain needs no trimming, and the discriminator adapts to each line and to drift by retraining digital weights alone.

\textbf{Cryogenic system models.} System studies model and optimize cryogenic control, readout and decoding at the system level~\cite{tannu2017taming,jokar2022digiq,byun2022xqsim,min2023qisim,holmes2020nisq,knapen2026pinball}, e.g., QIsim trims the RX's state-decision unit to scale the qubit count under a 4\,K budget~\cite{min2023qisim}. None optimizes readout for ML-based FDM discrimination or weighs its readout-fidelity cost against the larger code distance its power saving allows.

\section{Conclusion}\label{sec:conclusion}
In a cryo-CMOS QCI, a conventional readout RX BB spends its power not on discrimination but on the Nyquist-rate processing that its A/D boundary imposes. DIRAC redraws this boundary: the fixed, linear stages, demodulation and block averaging, are computed in analog, and only the down-sampled block averages are digitized by TDCs for the small, trainable MF and NN. Two co-designs make this practical: the readout TX BB supplies the square-wave LOs as the sign bits of its NCO words, removing per-tone PLLs, and a per-line calibration flow trains the MF jointly with the NN to absorb analog non-idealities, permitting low-bit-width digital processing at low voltages with FBB. Evaluated at 4\,K on 1.12\,M experimental shots, DIRAC matches FP32 fidelity and cuts the 4\,K QCI power per physical qubit by 50.4\% relative to FxP, which, at a fixed LQ count and 4\,K budget, allows a surface-code distance larger by 2 to 6 and a 1.6--7.3$\times$ lower LER on Willow-72.

\begin{acks}
We thank Prof. Boris Murmann from the University of Hawai\textquoteleft i for helpful discussions.
This research is supported by funding from the Quantum Research Institute at the University of Michigan.
We acknowledge the use of generative AI (Claude by Anthropic) in helping edit this manuscript.
\end{acks}

\bibliographystyle{ACM-Reference-Format-unsorted}
\bibliography{refs}

\end{document}